\documentclass[11pt, a4paper]{article}
\usepackage{amsmath,amssymb}
\usepackage[table]{xcolor}
\usepackage{arydshln}
\usepackage{graphicx,color,here}
\usepackage{ascmac}
\usepackage{xcolor}
\usepackage{ctable,dcolumn}
\usepackage{multirow}
\usepackage{listliketab}
\usepackage[numbers,authoryear]{natbib}
\bibpunct[,]{(}{)}{;}{a}{}{,}
\usepackage{setspace}
\usepackage{bm}
\def\vector#1{\mbox{\boldmath $#1$}}

\begin{document}
% \linenumbers

\title{A Bayesian approach to multi-task learning with network lasso}
\author{
  Kaito Shimamura
  \footnote{NTT Advanced Technology Corporation,
  Muza Kawasaki Central Tower, 1310 Omiya-cho, Saiwai-ku, Kawasaki-shi, Kanagawa 212-0014, Japan.
  Graduate School of Informatics and Engineering, The University of Electro-Communications, 1-5-1 Chofugaoka, Chofu-shi, Tokyo 182-8585, Japan.
  kaito.shimamura@ai.lab.uec.ac.jp (corresponding author)
  },\quad
  Shuichi Kawano
  \footnote{Graduate School of Informatics and Engineering, The University of Electro-Communications, 1-5-1 Chofugaoka, Chofu-shi, Tokyo 182-8585, Japan.
  skawano@ai.lab.uec.ac.jp
  }
}
\date{\empty}
\maketitle
\begin{abstract}
  Network lasso is a method for solving a multi-task learning problem through the regularized maximum likelihood method.
  A characteristic of network lasso is setting a different model for each sample.
  The relationships among the models are represented by relational coefficients.
  A crucial issue in network lasso is to provide appropriate values for these relational coefficients.
  In this paper, we propose a Bayesian approach to solve multi-task learning problems by network lasso.
  This approach allows us to objectively determine the relational coefficients by Bayesian estimation.
  The effectiveness of the proposed method is shown in a simulation study and a real data analysis.
\end{abstract}
\section{Introduction}

One important research topic is personalization, which is setting a different prediction model for each sample.
For example, in the field of personalized medicine, disease should not be predicted by a single model for all patients but rather predicted using a different model for each patient.
This issue is common to many research fields.
To achieve this aim, multi-task learning (\cite{evgeniou2007multi}; \cite{argyriou2008convex}; \cite{lengerich2018personalized}) is useful.

\cite{hallac2015network} proposed solving a multi-task learning problem through the regularized maximum likelihood method, which is also referred to as network lasso.
A characteristic of network lasso is to estimate models for each sample and group the models simultaneously.
Various applications of network lasso have been reported in the literature (e.g., \cite{ambos2018logistic}; \cite{jung2020duality}; \cite{tran2020classifying}).
The estimation of network lasso uses relational coefficients, which are quantitative expressions of the relationships among the samples.
However, no estimating methods for the relational coefficients in network lasso have yet been reported.

To overcome this problem, we propose a Bayesian framework for network lasso that includes estimating the relational coefficients.
We define the likelihood and the prior distribution of the regression coefficients, and consider network lasso from a Bayesian perspective.
Under this framework, we assume the Dirichlet prior distribution for the relational coefficients.
Models are estimated by the MCMC algorithm.
As a result, our method provides more accurate estimation than existing methods, even if the relational coefficients of some of the samples are unknown.
In addition, by using a prior distribution that induces variable selection, irrelevant features are excluded from the model.
This should allow us to expand the applicable fields of data compared to the existing method.

The remainder of the paper is organized as follows.
Section 2 describes regression modeling based on network lasso.
In Section 3, we propose a Bayesian approach to multi-task learning with network lasso.
Section 4 focuses on related work.
Section 5 compares the performance of the proposed method with the existing method by Monte Carlo simulation, and Section 6 compares the performance by real data analysis.
Concluding remarks are given in Section 7.

\section{Regression modeling based on network lasso}

Suppose that we have observed data $\{(y_i,\vector{x}_i);i=1,2,\cdots,n\}$ for response variable $y$ and $p$-dimensional predictor variables $\vector{x}=(x_1,x_2,\cdots,x_p)^T$.
A linear regression model is assumed as follows:
\begin{eqnarray}
  y_i=\vector{x}_i^T\vector{w}_i+\epsilon_i,
  \label{base_model}
\end{eqnarray}
where $\vector{w}_i=(w_{i1},w_{i2},\cdots,w_{ip})^T\ (i=1,2,\cdots,n)$ is a $p$-dimensional regression coefficient vector and $\epsilon_i$ is an error variable distributed as $\mbox{N}(0, \sigma^2)$, in which $\sigma^2\ (>0)$ is a variance parameter.
In a general regression problem, all regression coefficients are assumed to be the same (i.e., $\vector{w}_1 = \vector{w}_2 = \cdots = \vector{w}_n$).
In ($\ref{base_model}$), a different regression coefficient $\vector{w}_i$ is assumed for each sample.
This can be regarded as considering a different model for each sample.

For the overall model, we consider solving the regression problem by minimizing as follows:
\begin{eqnarray}
  \min_{\bm w_1,\cdots, \bm w_n}
  \sum_{i = 1}^n
  (y_i-\vector{x}_i^T\vector{w}_i)^2
  +
  \lambda
  \sum_{(i_1,i_2)\in\mathcal{E}}
  r_{i_1i_2}
  {\|\vector{w}_{i_1}-\vector{w}_{i_2} \|_2}
  ,
  \label{network_lasso_regression}
\end{eqnarray}
where $\lambda\ (>0)$ is a regularization parameter, $\mathcal{E}$ is a set whose elements are pairs of sample subscripts, and $r_{i_1i_2}$ is a relational coefficient.
In this minimization problem, the second term allows us to estimate $\hat{\vector{w}}_{i_1}=\hat{\vector{w}}_{i_2}$.
If $\hat{\vector{w}}_{i_1}=\hat{\vector{w}}_{i_2}$, then the $i_1$th sample and the $i_2$th sample belong to the same model.
The relational coefficient $r_{i_1i_2}$ is the strength of the relationship between samples $(y_{i_1},\vector{x}_{i_1})$ and $(y_{i_2},\vector{x}_{i_2})$.
The stronger the relationship is, the greater the value of $r_{i_1i_2}$ is.
\cite{hallac2015network} referred to the minimization problem in ($\ref{network_lasso_regression}$) as network lasso.

Here we focus on the relational coefficient $r_{i_1i_2}$.
The more correctly the relational coefficients are set, the better the accuracy of estimation of network lasso is.
In network lasso, the relational coefficients are determined independently of ($\ref{network_lasso_regression}$).
For example, if the samples have location information, then the relational coefficients are defined by the distances between the samples.
On the other hand, if the data do not contain such information, the determination of the relational coefficients becomes difficult.
Therefore, we propose an approach based on Bayesian sparse modeling.
Adopting this approach allows us to estimate the model and the relational coefficients simultaneously.

\section{Proposed method}

In this section, we introduce multi-task learning using network lasso from the viewpoint of Bayesian sparse modeling.
Bayesian sparse modeling, as represented by Bayesian lasso \citep{Park_2008}, has advantages such as quantifying posterior uncertainty and estimating hyperparameters.
The proposed method overcomes the problem described in the previous section.

\subsection{Model}

We consider a likelihood and a prior distribution of regression coefficients as follows.
\begin{eqnarray}
f(\vector{y}|X, W, \sigma^2)
&=&
(2\pi\sigma^2)^{-n/2}
\exp\left\{
-\frac{1}{2\sigma^2}\sum_{i=1}^n
(y_i-\vector{w}_i^T\vector{x}_i)^2
\right\}
,
\nonumber
\\
\pi(W|\lambda_1, \lambda_2, R, \sigma^2)
&\propto&
\prod_{(i_1,i_2)\in\mathcal{E}}
\frac{\lambda_1r_{i_1i_2}}{\sqrt{\sigma^2}}
\exp\left\{
-\frac{\lambda_1r_{i_1i_2}}{\sqrt{\sigma^2}}
\|\vector{w}_{i_1}-\vector{w}_{i_2}\|_2
\right\}
\label{prior}
\\
&&\quad\times
\prod_{i=1}^n
\left(
\frac{\lambda_2}{\sqrt{\sigma^2}}
\right)^p
\exp\left\{
-\frac{\lambda_2}{\sqrt{\sigma^2}}
\|\vector{w}_{i}\|_1
\right\}
\nonumber
,
\end{eqnarray}
where $\vector{y}=(y_1, y_2, \cdots, y_n)^T$ is the $n$-dimensional response vector, $\lambda_1\ (>0)$ and $\lambda_2\ (>0)$ are hyperparameters,
$X=(\vector{x}_1, \vector{x}_2, \cdots, \vector{x}_n)^T$ is the $n \times p$ design matrix, and $W=(\vector{w}_1, \vector{w}_2, \cdots, \vector{w}_n)^T$ is the $n \times p$ regression coefficient matrix.
The prior distribution (\ref{prior}) simultaneously plays a role in the grouping of individuals and the variable selection.
The hyperparameter $\lambda_1$ controls the degree of individual grouping and $\lambda_2$ controls the degree of variable selection.
Furthermore, we define an $n \times n$ relational matrix $R$ whose $(i_1,i_2)$-th element is the relational coefficient $r_{i_1i_2}$.
In particular, we assume that all diagonal components of $R$ are zero.

Next, we assume the following prior distribution for the relational coefficients:
\begin{eqnarray*}
  \{r_{i_1i_2}^{-1}\}
  &\sim&
  \mbox{Dir}(\alpha,\cdots,\alpha),
\end{eqnarray*}
where $\mbox{Dir}(\cdot)$ represents the Dirichlet distribution and $\alpha\ (>0)$ is a hyperparameter.
Note that we could also assume independent gamma distributions for $r_{i_1i_2}^{-1}$.
However, assuming independent prior distributions for parameters with relative measures, such as relational coefficients, empirically reduces the predictive accuracy of the model.
Therefore, we use the Dirichlet distribution, which takes the joint distribution of relational coefficients into account.
We also assume the following prior distribution for $\lambda_1$:
\begin{eqnarray*}
  \lambda_1^{-1}
  &\sim&
  \mbox{Ga}(\alpha\#\mathcal{E}, 1/2)
  ,
\end{eqnarray*}
where $\#\mathcal{E}$ is the number of elements in the set $\mathcal{E}$ and $\mbox{Ga}(\cdot,\cdot)$ represents the gamma distribution.

Thereby, the tuning parameters in the proposed method are $\alpha$ and $\lambda_2$.
The $\alpha$ controls a group of individuals.
When $\alpha$ is small, $\lambda_1$ is large and the variance of $\{r_{i_1i_2}\}$ is large.
In other words, the smaller $\alpha$ is, the more $\|\vector{w}_{i_1}-\vector{w}_{i_2}\|_2$ terms are estimated to be zero.

\subsection{Estimation by MCMC}

We use Gibbs sampling to estimate the proposed model.
Since it is difficult to obtain the full conditional distributions from the prior distribution in $(\ref{prior})$, we derive them by rewriting the prior distribution.
Using scale mixtures of normal distributions \citep{andrews1974scale}, we obtain the following transformation:
\begin{eqnarray*}
  \pi(W, \{r_{i_1i_2}\}, \lambda_1, \sigma^2| \alpha, \lambda_2)
  &\propto&
  \prod_{(i_1,i_2)\in\mathcal{E}}
  \frac{\lambda_1r_{i_1i_2}}{\sqrt{\sigma^2}}
  \exp\left\{
    -\frac{\lambda_1r_{i_1i_2}}{\sqrt{\sigma^2}}
    \|\vector{w}_{i_1}-\vector{w}_{i_2}\|_2
  \right\}
  \\
  &&\times
  \prod_{i=1}^n \prod_{j=1}^p
  \frac{\lambda_2}{\sqrt{\sigma^2}}
  \exp\left\{
    -\frac{\lambda_2}{\sqrt{\sigma^2}}
    |w_{ij}|
  \right\}
  \\
  &&\times
  \lambda_1^{-(\alpha\#\mathcal{E}-1)}
  \exp\left\{
    -\frac{1}{2\lambda_1}
  \right\}
  \pi(\sigma^2)
  \prod_{(i_1, i_2)\in\mathcal{E}} r_{i_1i_2}^{-(\alpha-1)}
  \\
  &\propto&
  \prod_{(i_1,i_2)\in\mathcal{E}}
  \frac{\lambda_1r_{i_1i_2}}{\sqrt{\sigma^2}}
  \int_{0}^{\infty}
  g_1(\vector{w}_{i_1},\vector{w}_{i_2},r_{i_1i_2},\lambda_1,\tau_{i_1i_2},\sigma^2)
  d\tau_{i_1i_2}
  \\
  &&\times
  \prod_{i=1}^n\prod_{j=1}^p
  \frac{\lambda_2^2}{\sqrt{\sigma^2}}
  \int_{0}^{\infty}
  g_2(w_{ij},\widetilde{\tau}_{ij},\sigma^2,\lambda_2)
  d\widetilde{\tau}_{ij}
  \\
  &&\times
  \lambda_1^{-(\alpha\#\mathcal{E}-1)}
  \exp\left\{
    -\frac{1}{2\lambda_1}
  \right\}
  \pi(\sigma^2)
  \prod_{(i_1, i_2)\in\mathcal{E}} r_{i_1i_2}^{-(\alpha-1)}
  ,
\end{eqnarray*}
where
\begin{eqnarray*}
  g_1(\vector{w}_{i_1},\vector{w}_{i_2},r_{i_1i_2},\lambda_1,\tau_{i_1i_2},\sigma^2)
  &=&
  \tau_{i_1i_2}^{-1/2}
  \exp\left\{
    -\frac{\lambda_1^2r_{i_1i_2}^2}{2\tau_{i_1i_2}\sigma^2}
    \|\vector{w}_{i_1}-\vector{w}_{i_2}\|_2^2
  \right\}
  \exp\left\{
    -\frac{1}{2}\tau_{i_1i_2}
  \right\}
  ,
  \\
  g_2(w_{ij},\widetilde{\tau}_{ij},\sigma^2,\lambda_2)
  &=&
  \widetilde{\tau}_{ij}^{-1/2}
  \exp\left\{
    -\frac{1}{2\widetilde{\tau}_{ij}\sigma^2}
    w_{ij}^2
  \right\}
  \exp\left\{
    -\frac{\lambda_2^2}{2}\widetilde{\tau}_{ij}
  \right\}
  .
\end{eqnarray*}
This transformation allows us to rewrite the prior distribution as follows:
\begin{eqnarray}
  \pi(W|\{1/r_{i_1i_2}\}, 1/\lambda_1, \{\tau_{i_1i_2}^{-1}\}, \{\widetilde{\tau}_{ij}^{-1}\}, \sigma^2)
  &\propto&
  \prod_{(i_1,i_2)\in\mathcal{E}}
  \mbox{N}(\|\vector{w}_{i_1}-\vector{w}_{i_2}\|_2\ |\ 0, \tau_{i_1i_2}\sigma^2/\lambda_1^2r_{i_1i_2}^2)
  \nonumber
  \\
  &&\times
  \prod_{i=1}^n\prod_{j=1}^p
  \mbox{N}(w_{ij}\ |\ 0, \widetilde{\tau}_{ij}\sigma^2)
  ,
  \label{prior_distribution}
  \\
  \{r_{i_1i_2}^{-1}\}
  &\sim&
  \mbox{Dir}(\alpha,\cdots,\alpha)
  ,
  \nonumber
  \\
  \lambda_1^{-1}
  &\sim&
  \mbox{Ga}(\alpha\#\mathcal{E}, 1/2)
  ,
  \nonumber
  \\
  \tau_{i_1i_2}
  &\sim&
  \mbox{Exp}(1/2)
  ,
  \nonumber
  \\
  \widetilde{\tau}_{ij}
  &\sim&
  \mbox{Exp}(\lambda_2^2/2)
  ,
  \nonumber
  \\
  \sigma^2
  &\sim&
  \mbox{IG}(\nu_0/2,\eta_0/2)
  ,
  \nonumber
\end{eqnarray}
where $\mbox{Exp}(\cdot)$ represents the exponential distribution and $\mbox{IG}(\cdot,\cdot)$ represents the inverse-gamma distribution.
The inverse-gamma probability density function is given by
\begin{eqnarray}
  \mbox{IG}(x|\nu,\eta)
  =
  \frac{\eta^\nu}{\Gamma(\nu)
  x^{-(\nu+1)}}
  \exp\left(
  -\frac{\eta}{x}
  \right)
  ,
  \label{inverse_gamma}
\end{eqnarray}
where $\nu\ (>0)$ is a shape parameter, $\eta\ (>0)$ is a scale parameter, and $\Gamma(\cdot)$ is the Gamma function.
This allows us to consider $\tau_{i_1i_2}$ and $\widetilde{\tau}_{ij}$ as new parameters in addition to $W$, $\{ r_{i_1i_2} \}$, $ \lambda_1$, and $\sigma^2$.
By using these prior distributions, the posterior distribution can be estimated by Gibbs sampling.
The details of the hierarchical representation and the sampling algorithm are given in \ref{gibbs_sammpling}.

We note that the prior distribution in (\ref{prior_distribution}) is identical to the Dirichlet--Laplace prior distribution \citep{bhattacharya2015dirichlet}.
The prior distributions in (\ref{prior_distribution}) are expressed as follows:
\begin{eqnarray*}
  \pi(W|\sigma^2)
  &\propto&
  \prod_{(i_1,i_2)\in\mathcal{E}}
  \frac{1}{\sqrt{\sigma^2}}
  \mbox{DL}(\|\vector{w}_{i_1}-\vector{w}_{i_2}\|_2/\sqrt{\sigma^2}\ |\ \alpha)
  \\
  &&\times
  \prod_{i=1}^n
  \prod_{j=1}^p
  \frac{1}{\sqrt{\sigma^2}}
  \mbox{Laplace}(w_{ij}/\sqrt{\sigma^2}\ |\ \lambda_2),
  \\
  \sigma^2
  &\sim&
  \mbox{IG}(\nu_0/2,\eta_0/2)
  ,
\end{eqnarray*}
where $\mbox{DL}(\cdot)$ represents the Dirichlet--Laplace prior distribution.
Based on this expression, the proposed model can be regarded as applying the Dirichlet--Laplace prior to the grouping of individuals and the Laplace distribution to the selection of variables.
The details of the Dirichlet--Laplace prior distribution are described in \ref{dirichlet_laplace_prior}.

The Dirichlet--Laplace prior distribution is a type of global--local prior distribution \citep{polson2010shrink}.
The important features of the global--local prior distribution include that it has a peak at the origin and that it has heavy tails.
These features allow us to perform more robust estimation than using the Laplace distribution.
Various global--local priors have been proposed: e.g., the normal-exponential-gamma distribution \citep{griffin2005alternative}, the normal-gamma distribution \citep{griffin2010inference}, and the horseshoe distribution \citep{carvalho2010horseshoe}.

\section{Relationship with peripheral research}

We here summarize the research peripherally related to the proposed method.
First, we describe localized lasso \citep{yamada2017localized}, which performs grouping of individuals and variable selection simultaneously.
Next, we describe another Bayesian approach for network lasso.

\subsection{Localized lasso}

The localized lasso proposed by \cite{yamada2017localized} considers the following minimization problem:
\begin{eqnarray*}
  \min_{\bm w_1,\cdots, \bm w_n}
  \sum_{i=1}^n (y_i-\vector{x}_i^T\vector{w}_i)^2
  +
  \lambda_1 \sum_{(i_1,i_2)\in\mathcal{E}}
  r_{i_1i_2}
  {\|\vector{w}_{i_1}-\vector{w}_{i_2} \|_2}
  +
  \lambda_2 \sum_{i=1}^n\|\vector{w}_i\|_1^2
  .
\end{eqnarray*}
This minimization is based on network lasso and performs grouping of individuals and variable selection.
In addition, \cite{yamada2017localized} proposed an efficient iterative algorithm for localized lasso. This algorithm converges to the global optimal solution without the need for tuning the parameters $\lambda_1$ and $\lambda_2$.

Our proposed method uses the $L_1$ norm, while localized lasso uses the $L_{1,2}$ norm.
Because the $L_{1,2}$ norm is used, at least one of the regression coefficients for a sample is always non-zero for localized lasso.
Since our proposed method employs the $L_1$ norm, $\vector{w}_i = \vector{0}$ may arise.
To avoid this case, we adjust the hyperparameter $\lambda_2$ to obtain $\vector{w}_i \neq \vector{0}$.

Localized lasso requires knowledge of the relational coefficients.
That is, if the relational coefficients are unknown, localized lasso needs to estimate relational coefficients from data.
This means that the performance of localized lasso depends on the method for determining relational coefficients.

\subsection{Network lasso in a Bayesian framework}

\cite{rad2017robust} proposed using the following likelihood and prior distribution:
\begin{eqnarray}
f(\vector{y}|X, W, \sigma^2)
&=&
(2\pi\sigma^2)^{-n/2}
\exp\left\{
-\frac{1}{2\sigma^2}\sum_{i=1}^n
(y_i-\vector{w}_i^T\vector{x}_i)^2
\right\}
,
\label{bnl}
\\
\pi(W|\lambda, \sigma^2)
&\propto&
\prod_{(i_1,i_2)\in\mathcal{E}}
\frac{\lambda}{\sqrt{\sigma^2}}
\exp\left\{
-\frac{\lambda}{\sqrt{\sigma^2}}
\|\vector{w}_{i_1}-\vector{w}_{i_2}\|_2
\right\}
\nonumber
.
\end{eqnarray}
Unlike the prior distribution in ($\ref{prior_distribution}$), this prior distribution does not consider any relational coefficients $r_{i_1 i_2}$.
Moreover, this prior does not allow variable selection.
Gibbs sampling is used to estimate the posterior distribution.
\cite{rad2017robust} pointed out that the MAP estimates of regression coefficients given by (\ref{bnl}) are similar to those from network lasso.

\cite{rad2017robust} showed that ADMM allows for fast computation of estimates of regression coefficients.
On the other hand, \cite{rad2017robust} pointed out that using ADMM sacrifices the quantification of the posterior uncertainty.
By using the quantification of the posterior uncertainty, it is possible to evaluate the estimates from the shape of the posterior distribution, rather than focusing only on the results of point estimation.

\cite{rad2017robust} set assumptions as the target data being an evenly spaced time series or having lattice position information, and determined $\mathcal{E}$ based on these assumptions.
Therefore, the method can be applied only to data that satisfy the above prerequisites.

\section{Simulation study \label{simulationStudy}}

We demonstrated our proposed method with artificial data.
We simulated data $(y_i, \vector{x}_i) \ (i=1,2,\cdots,n)$ from the following model:
\begin{eqnarray*}
  y_i=\vector{x}_i^T\vector{w}_{k,i}^*+\epsilon_i,
  \qquad
  (i=1,2,\cdots,n),
\end{eqnarray*}
where $\vector{w}^*_{k,i}$ is a vector of the $p$-dimensional true regression coefficient and $\epsilon_i$ is an error distributed as $\mbox{N}(0, 1)$.
In addition, $\vector{x}_i\ (i=1,2,\cdots,n)$ is generated from a uniform distribution on $[-1,1]$.
We considered several settings: $n=30,\ 50$ and $p=5,\ 10$.
When $p=10$, the features consist of five irrelevant features and five relevant features.
The data have three groups, and the sample size for each group is the same.
The true regression coefficients were set as follows:
\begin{itemize}
  \item When $p=5$,
  \begin{eqnarray*}
    \vector{w}_{1,i}^* &=& (5.0,\ 1.0,\ -1.0,\ 0.0,\ 0.0)^T, \\
    \vector{w}_{2,i}^* &=& (0.0,\ 1.0,\ -5.0,\ 1.0,\ 0.0)^T, \\
    \vector{w}_{3,i}^* &=& (0.0,\ 0.0,\ 0.0,\ 0.5,\ -0.5)^T
    .
  \end{eqnarray*}
  \item When $p=10$,
  \begin{eqnarray*}
    \vector{w}_{1,i}^* &=& (5.0,\ 1.0,\ -1.0,\ 0.0,\ 0.0,\ \vector{0}_{5}^T)^T, \\
    \vector{w}_{2,i}^* &=& (0.0,\ 1.0,\ -5.0,\ 1.0,\ 0.0,\ \vector{0}_{5}^T)^T, \\
    \vector{w}_{3,i}^* &=& (0.0,\ 0.0,\ 0.0,\ 0.5,\ -0.5,\ \vector{0}_{5}^T)^T
    .
  \end{eqnarray*}
\end{itemize}
The subscript set $\mathcal{E}$ was also generated by random numbers.
Let $\mathcal{F}$ be the set of subscript pairs $(i, j)$ such that $\vector{x}_i$ and $\vector{x}_j$ belong to the same group and let $\mathcal{G}$ be the set of subscript pair $(i,j)$ such that $\vector{x}_i$ and $\vector{x}_j$ belong to the different groups.
The subscript set $\mathcal{E}$ was given by $\mathcal{E} =\{ f, g \}$, where $f \subset \mathcal{F}$ and $g \subset \mathcal{G}$.
We assume that $\#f = \mbox{TR} \times \#\mathcal{F}$ and $\#g = \mbox{FR} \times \#\mathcal{G}$, where TR and FR are constants between zero and one.

As the estimation accuracy, we used the MSE given by
\begin{eqnarray*}
  \mbox{MSE}
  &=&
  \mbox{E}
  \left[
    (\vector{y}^*-\hat{\vector{y}})^T(\vector{y}^*-\hat{\vector{y}})
  \right]\\
  &=&
  \sum_{i=1}^n (\vector{w}_{k,i}^*-\hat{\vector{w}}_i)^{T}
  \Sigma_i
  (\vector{w}_{k,i}^*-\hat{\vector{w}}_i)
  ,
\end{eqnarray*}
where
\begin{eqnarray*}
&&
\vector{y}^*=(y^*_1,y^*_2,\cdots,y^*_n)^T,
\quad
y^*_i=\vector{w}_{k,i}^{*T}\vector{x}_i,
\\
&&
\hat{\vector{y}}=(\hat{y}_1,\hat{y}_2,\cdots,\hat{y}_n)^T,
\quad
\hat{y}_i=\hat{\vector{w}}_i^{T}\vector{x}_i,
\end{eqnarray*}
$\hat{\vector{w}}_i$ is an estimate of $\vector{w}_i$, and $\Sigma_i$ is the variance-covariance matrix of $\vector{x}_i$.

The dataset was generated 50 times.
We computed the mean and standard deviation of MSE from the 50 repetitions.
For each generated dataset, the estimates were obtained using 50,000 iterations of a Gibbs sampler.
Candidates of the hyperparameters were set as $10^{l-i}\ (i = 1,2, \cdots m)$.
In order to make a fair comparison, the results were compared using the hyperparameter that maximized the MSE.

The existing method to be compared was localized lasso (LL).
We also compared a method that assumes fixed values for $r_{i_1i_2}$ and $\lambda_1$ instead of assuming prior distributions.
We refer to this method as the Bayesian approach to multi-task learning with network lasso (BMN).
The proposed method is referred to as the Bayesian approach to multi-task learning with network lasso using the Dirichlet--Laplace distribution (DLBMN).
By comparing BMN and DLBMN, we can consider whether assuming a prior distribution for the hyperparameters increases the accuracy of the estimation.

Tables \ref{simulationResults1} to \ref{simulationResults4} show the means and standard deviations of the MSEs.
Not surprisingly, all tables show that the larger the TR is and the smaller the FR is, the smaller the MSE is.
Also, for all tables, regardless of TR, the increase in MSE with increasing FR is smaller for DLBMN than for LL and BMN.
From the results for $n=30$ and $p=5$ in Table \ref{simulationResults1}, when $\mbox{TR}=0.2$, the MSE is large regardless of the TF.
The results for $n=30$ and $p=10$ in Table \ref{simulationResults2} also show that the MSE is large regardless of TF for $\mbox{TR}=0.2$.
When FR is greater than $0.1$, LL and BMN have MSEs greater than 20, while the MSE of DLBMN is very small.
In the results for $n=120$ in Table \ref{simulationResults3} and Table \ref{simulationResults4}, the MSE is small when $\mbox{TR}=0.2$ and $\mbox{FR}=0.01$, unlike the results for $n=30$.
In the results for $n=120$ and $p=5$, unlike the other results, DLBMN has a small MSE even with $\mbox{TR}=1.0$ and $\mbox{FR}=0.4$.
DLBMN has a smaller MSE than LL and BMN in most of the results.
In particular, the number of results such that the MSE was less than $1.0$ was 34 for DLBMN, compared to 19 for LL and 13 for BMN.

\begin{table}[H]
  \begin{center}
    \caption{Means and standard deviations of MSEs in simulation study with $n=30$ and $p=5$.}
    \label{simulationResults1}
    \begin{tabular}{ccrlrlrl}
      \cline{3-8}
       & & \multicolumn{2}{c}{LL} & \multicolumn{2}{c}{BMN} & \multicolumn{2}{c}{DLBMN} \\
      \hline
       \multirow{6}{*}{TR=0.20} & FR=0.01 &
       30.96 & [17.85] & 31.21 & [16.05] & 29.39 & [16.11] \\
       & FR=0.10 &
       89.26 & [21.85] & 82.29 & [17.33] & 79.13 & [18.66] \\
       & FR=0.20 &
       102.52 & [13.90] & 97.41 & [13.55] & 95.83 & [14.73] \\
       & FR=0.30 &
       105.98 & [11.14] & 101.82 & [12.01] & 99.85 & [13.13] \\
       & FR=0.40 &
       108.18 & [11.24] & 104.91 & [11.40] & 102.96 & [12.60] \\
       & FR=1.00 &
       109.36 & [10.71] & 107.36 & [11.49] & 105.30 & [11.36] \\
       \cline{1-8}
       \multirow{6}{*}{TR=0.40} & FR=0.01 &
       1.99 & [3.29] & 2.25 & [3.45] & 2.10 & [3.41] \\
       & FR=0.10 &
       22.36 & [14.65] & 40.98 & [18.30] & 14.34 & [16.57] \\
       & FR=0.20 &
       67.87 & [17.34] & 74.74 & [18.86] & 68.57 & [24.93] \\
       & FR=0.30 &
       85.66 & [15.42] & 86.50 & [16.06] & 84.29 & [17.67] \\
       & FR=0.40 &
       94.31 & [14.44] & 93.38 & [14.67] & 91.41 & [15.14] \\
       & FR=1.00 &
       106.79 & [11.50] & 103.66 & [12.05] & 102.14 & [11.81] \\
       \cline{1-8}
       \multirow{6}{*}{TR=0.60} & FR=0.01 &
       0.31 & [0.18] & 0.36 & [0.28] & 0.38 & [0.26] \\
       & FR=0.10 &
       3.37 & [3.94] & 16.31 & [13.09] & 0.88 & [1.48] \\
       & FR=0.20 &
       34.01 & [21.75] & 53.62 & [20.83] & 22.31 & [30.43] \\
       & FR=0.30 &
       64.28 & [19.06] & 72.42 & [18.57] & 57.60 & [34.38] \\
       & FR=0.40 &
       81.65 & [16.92] & 82.68 & [17.09] & 80.49 & [20.15] \\
       & FR=1.00 &
       103.89 & [11.90] & 100.75 & [13.26] & 99.90 & [13.33] \\
       \cline{1-8}
       \multirow{6}{*}{TR=0.80} & FR=0.01 &
       0.28 & [0.16] & 0.30 & [0.20] & 0.32 & [0.22] \\
       & FR=0.10 &
       0.78 & [0.75] & 5.25 & [6.01] & 0.38 & [0.27] \\
       & FR=0.20 &
       12.51 & [11.42] & 32.57 & [18.45] & 3.35 & [9.70] \\
       & FR=0.30 &
       39.92 & [21.39] & 58.10 & [20.97] & 23.20 & [33.60] \\
       & FR=0.40 &
       65.45 & [19.54] & 72.60 & [18.84] & 61.29 & [36.77] \\
       & FR=1.00 &
       101.05 & [13.13] & 96.81 & [14.36] & 96.90 & [13.88] \\
       \cline{1-8}
       \multirow{6}{*}{TR=1.00} & FR=0.01 &
       0.28 & [0.16] & 0.30 & [0.21] & 0.28 & [0.19] \\
       & FR=0.10 &
       0.46 & [0.29] & 2.56 & [3.76] & 0.31 & [0.20] \\
       & FR=0.20 &
       3.96 & [5.29] & 18.07 & [15.71] & 0.36 & [0.23] \\
       & FR=0.30 &
       19.81 & [17.85] & 40.87 & [21.39] & 10.65 & [25.81] \\
       & FR=0.40 &
       43.95 & [21.58] & 59.26 & [20.16] & 36.48 & [37.36] \\
       & FR=1.00 &
       96.15 & [14.20] & 92.77 & [14.24] & 94.04 & [14.58] \\
       \hline
    \end{tabular}
  \end{center}
\end{table}

\begin{table}[H]
  \begin{center}
    \caption{Means and standard deviations of MSEs in simulation study with $n=30$ and $p=10$.}
    \label{simulationResults2}
    \begin{tabular}{ccrlrlrl}
       \cline{3-8}
       & & \multicolumn{2}{c}{LL} & \multicolumn{2}{c}{BMN} & \multicolumn{2}{c}{DLBMN} \\
       \hline
       \multirow{6}{*}{TR=0.20} & FR=0.01 &
       85.60 & [29.38] & 53.85 & [21.58] & 51.84 & [19.60] \\
       & FR=0.10 &
       130.96 & [17.11] & 110.48 & [18.36] & 110.13 & [19.39] \\
       & FR=0.20 &
       131.11 & [13.38] & 120.50 & [14.54] & 118.54 & [15.13] \\
       & FR=0.30 &
       130.16 & [13.31] & 123.04 & [12.56] & 120.74 & [13.29] \\
       & FR=0.40 &
       129.58 & [12.84] & 124.71 & [12.05] & 122.72 & [12.92] \\
       & FR=1.00 &
       129.35 & [13.34] & 124.70 & [12.55] & 122.94 & [12.69] \\
       \cline{1-8}
       \multirow{6}{*}{TR=0.40} & FR=0.01 &
       20.17 & [21.40] & 8.32 & [8.67] & 5.31 & [6.81] \\
       & FR=0.10 &
       93.14 & [27.10] & 79.07 & [19.33] & 60.55 & [26.00] \\
       & FR=0.20 &
       122.32 & [17.46] & 108.41 & [15.80] & 104.73 & [17.28] \\
       & FR=0.30 &
       126.18 & [15.40] & 116.53 & [14.15] & 114.26 & [14.94] \\
       & FR=0.40 &
       127.55 & [14.47] & 119.65 & [12.89] & 117.21 & [13.87] \\
       & FR=1.00 &
       129.21 & [13.50] & 123.42 & [12.53] & 122.49 & [13.44] \\
       \cline{1-8}
       \multirow{6}{*}{TR=0.60} & FR=0.01 &
       12.02 & [15.26] & 2.35 & [4.91] & 1.74 & [3.80] \\
       & FR=0.10 &
       48.17 & [24.13] & 53.31 & [21.78] & 12.69 & [16.04] \\
       & FR=0.20 &
       104.08 & [23.43] & 94.27 & [18.32] & 75.69 & [32.33] \\
       & FR=0.30 &
       121.88 & [17.22] & 108.57 & [15.08] & 104.34 & [17.67] \\
       & FR=0.40 &
       126.76 & [15.00] & 114.92 & [13.54] & 112.10 & [15.23] \\
       & FR=1.00 &
       128.86 & [13.59] & 121.33 & [12.72] & 120.82 & [13.74] \\
       \cline{1-8}
       \multirow{6}{*}{TR=0.80} & FR=0.01 &
       9.33 & [14.75] & 1.87 & [5.95] & 1.33 & [3.06] \\
       & FR=0.10 &
       28.75 & [20.15] & 33.04 & [20.23] & 3.98 & [8.67] \\
       & FR=0.20 &
       77.24 & [29.98] & 77.41 & [21.23] & 30.33 & [32.15] \\
       & FR=0.30 &
       111.00 & [21.73] & 98.60 & [17.72] & 86.84 & [34.78] \\
       & FR=0.40 &
       122.71 & [16.16] & 108.14 & [15.38] & 107.07 & [16.76] \\
       & FR=1.00 &
       128.69 & [13.35] & 119.45 & [13.19] & 119.67 & [13.73] \\
       \cline{1-8}
       \multirow{6}{*}{TR=1.00} & FR=0.01 &
       7.93 & [8.08] & 1.43 & [2.08] & 0.78 & [0.88] \\
       & FR=0.10 &
       21.34 & [13.84] & 21.67 & [15.33] & 2.10 & [4.32] \\
       & FR=0.20 &
       57.37 & [29.16] & 59.66 & [23.64] & 10.07 & [15.18] \\
       & FR=0.30 &
       94.69 & [26.93] & 86.22 & [20.60] & 52.09 & [42.84] \\
       & FR=0.40 &
       116.11 & [20.41] & 100.84 & [18.46] & 101.56 & [23.35] \\
       & FR=1.00 &
       128.33 & [13.74] & 116.58 & [14.23] & 118.79 & [14.50] \\
       \hline
    \end{tabular}
  \end{center}
\end{table}

\begin{table}[H]
  \begin{center}
    \caption{Means and standard deviations of MSEs in simulation study with $n=120$ and $p=5$.}
    \label{simulationResults3}
    \begin{tabular}{ccrlrlrl}
       \cline{3-8}
       & &
       \multicolumn{2}{c}{LL} & \multicolumn{2}{c}{BMN} & \multicolumn{2}{c}{DLBMN} \\
       \hline
       \multirow{6}{*}{TR=0.20} & FR=0.01 &
       0.93 & [2.99] & 1.03 & [2.89] & 0.77 & [2.84] \\
       & FR=0.10 &
       201.14 & [32.03] & 222.27 & [27.69] & 33.10 & [64.60] \\
       & FR=0.20 &
       318.97 & [19.16] & 319.84 & [19.31] & 320.61 & [19.48] \\
       & FR=0.30 &
       348.21 & [16.70] & 347.98 & [17.16] & 347.75 & [17.00] \\
       & FR=0.40 &
       361.51 & [15.99] & 360.85 & [17.12] & 360.53 & [16.47] \\
       & FR=1.00 &
       380.77 & [14.87] & 381.24 & [15.29] & 380.17 & [15.98] \\
       \cline{1-8}
       \multirow{6}{*}{TR=0.40} & FR=0.01 &
       0.22 & [0.09] & 0.24 & [0.11] & 0.17 & [0.08] \\
       & FR=0.10 &
       2.53 & [1.38] & 13.22 & [9.58] & 0.31 & [0.12] \\
       & FR=0.20 &
       200.63 & [35.25] & 220.72 & [31.88] & 6.21 & [37.70] \\
       & FR=0.30 &
       290.24 & [21.22] & 293.01 & [21.31] & 301.01 & [21.26] \\
       & FR=0.40 &
       321.41 & [17.83] & 321.51 & [18.53] & 326.63 & [18.65] \\
       & FR=1.00 &
       367.69 & [15.22] & 367.48 & [15.63] & 366.98 & [16.57] \\
       \cline{1-8}
       \multirow{6}{*}{TR=0.60} & FR=0.01 &
       0.18 & [0.08] & 0.21 & [0.10] & 0.23 & [0.09] \\
       & FR=0.10 &
       0.86 & [0.26] & 1.53 & [0.53] & 0.21 & [0.09] \\
       & FR=0.20 &
       22.10 & [21.29] & 61.46 & [35.50] & 0.25 & [0.13] \\
       & FR=0.30 &
       200.57 & [36.42] & 218.47 & [32.46] & 11.02 & [52.97] \\
       & FR=0.40 &
       273.53 & [23.65] & 278.57 & [22.96] & 288.15 & [47.29] \\
       & FR=1.00 &
       353.08 & [16.13] & 353.43 & [16.33] & 355.18 & [17.32] \\
       \cline{1-8}
       \multirow{6}{*}{TR=0.80} & FR=0.01 &
       0.17 & [0.08] & 0.21 & [0.11] & 0.21 & [0.09] \\
       & FR=0.10 &
       0.61 & [0.18] & 0.78 & [0.23] & 0.19 & [0.08] \\
       & FR=0.20 &
       2.17 & [0.96] & 6.55 & [4.56] & 0.20 & [0.10] \\
       & FR=0.30 &
       60.41 & [38.97] & 102.47 & [43.17] & 0.25 & [0.14] \\
       & FR=0.40 &
       198.08 & [37.27] & 215.65 & [33.84] & 44.49 & [99.13] \\
       & FR=1.00 &
       338.08 & [17.27] & 339.14 & [17.21] & 344.62 & [18.09] \\
       \cline{1-8}
       \multirow{6}{*}{TR=1.00} & FR=0.01 &
       0.17 & [0.08] & 0.21 & [0.11] & 0.21 & [0.11] \\
       & FR=0.10 &
       0.50 & [0.15] & 0.55 & [0.17] & 0.18 & [0.08] \\
       & FR=0.20 &
       1.13 & [0.38] & 2.03 & [0.77] & 0.20 & [0.10] \\
       & FR=0.30 &
       5.91 & [4.57] & 20.82 & [16.34] & 0.23 & [0.12] \\
       & FR=0.40 &
       89.68 & [42.72] & 126.74 & [43.54] & 0.28 & [0.16] \\
       & FR=1.00 &
       322.05 & [18.44] & 324.17 & [18.42] & 335.27 & [18.19] \\
       \hline
     \end{tabular}
   \end{center}
 \end{table}

\begin{table}[H]
  \begin{center}
    \caption{Means and standard deviations of MSEs in simulation study with $n=120$ and $p=10$.}
    \label{simulationResults4}
    \begin{tabular}{ccrlrlrl}
      \cline{3-8}
       & &
       \multicolumn{2}{c}{LL} & \multicolumn{2}{c}{BMN} & \multicolumn{2}{c}{DLBMN} \\
       \hline
       \multirow{6}{*}{TR=0.20} & FR=0.01 &
       0.87 & [1.35] & 1.10 & [1.51] & 0.60 & [1.09] \\
       & FR=0.10 &
       361.14 & [26.82] & 353.15 & [22.02] & 355.39 & [22.64] \\
       & FR=0.20 &
       411.92 & [22.00] & 404.21 & [19.47] & 402.18 & [19.61] \\
       & FR=0.30 &
       425.32 & [20.88] & 420.40 & [19.48] & 416.43 & [19.86] \\
       & FR=0.40 &
       431.38 & [20.26] & 427.08 & [18.53] & 423.01 & [19.69] \\
       & FR=1.00 &
       438.29 & [18.46] & 430.65 & [16.93] & 431.33 & [18.02] \\
       \cline{1-8}
       \multirow{6}{*}{TR=0.40} & FR=0.01 &
       0.45 & [0.15] & 0.38 & [0.20] & 0.33 & [0.11] \\
       & FR=0.10 &
       126.75 & [60.35] & 177.94 & [48.44] & 0.41 & [0.16] \\
       & FR=0.20 &
       362.05 & [26.46] & 359.88 & [22.05] & 365.44 & [21.27] \\
       & FR=0.30 &
       396.84 & [22.85] & 391.57 & [18.87] & 394.19 & [19.66] \\
       & FR=0.40 &
       410.57 & [22.19] & 403.36 & [18.37] & 405.67 & [19.89] \\
       & FR=1.00 &
       432.03 & [19.76] & 423.16 & [16.56] & 424.74 & [18.32] \\
       \cline{1-8}
       \multirow{6}{*}{TR=0.60} & FR=0.01 &
       0.40 & [0.13] & 0.34 & [0.15] & 0.34 & [0.11] \\
       & FR=0.10 &
       4.32 & [3.59] & 15.55 & [11.60] & 0.37 & [0.13] \\
       & FR=0.20 &
       276.53 & [40.62] & 284.20 & [31.54] & 7.57 & [49.54] \\
       & FR=0.30 &
       363.08 & [25.91] & 356.69 & [21.10] & 371.72 & [21.68] \\
       & FR=0.40 &
       390.34 & [23.20] & 380.98 & [18.92] & 391.74 & [20.32] \\
       & FR=0.40 &
       426.53 & [20.83] & 416.03 & [16.99] & 419.09 & [18.76] \\
       \cline{1-8}
       \multirow{6}{*}{TR=0.80} & FR=0.01 &
       0.38 & [0.13] & 0.32 & [0.12] & 0.34 & [0.10] \\
       & FR=0.10 &
       1.47 & [0.69] & 2.54 & [1.26] & 0.37 & [0.12] \\
       & FR=0.20 &
       112.35 & [58.22] & 150.19 & [47.83] & 0.43 & [0.17] \\
       & FR=0.30 &
       311.25 & [33.57] & 309.14 & [27.34] & 198.92 & [176.32] \\
       & FR=0.40 &
       363.11 & [25.56] & 355.28 & [21.15] & 377.08 & [21.96] \\
       & FR=1.00 &
       419.47 & [21.53] & 409.90 & [17.67] & 415.17 & [19.82] \\
       \cline{1-8}
       \multirow{6}{*}{TR=1.00} & FR=0.01 &
       0.37 & [0.13] & 0.31 & [0.11] & 0.34 & [0.10] \\
       & FR=0.10 &
       0.99 & [0.43] & 1.17 & [0.44] & 0.36 & [0.11] \\
       & FR=0.20 &
       17.18 & [17.23] & 36.70 & [26.78] & 0.41 & [0.13] \\
       & FR=0.30 &
       227.63 & [49.11] & 237.62 & [38.73] & 15.23 & [72.67] \\
       & FR=0.40 &
       326.41 & [29.59] & 321.66 & [23.86] & 332.25 & [100.64] \\
       & FR=1.00 &
       411.29 & [21.01] & 401.90 & [16.85] & 410.47 & [18.92] \\
       \hline
    \end{tabular}
  \end{center}
\end{table}

\section{Application}

We applied our proposed method to a real dataset comprising Greater Sacramento area data (http://support.spatialkey.com/spatialkey-sample-csv-data/).
This dataset was used by \cite{hallac2015network}.
This dataset is a listing of real estate transactions for one week of May 2008.
It contains 985 samples of sales information, including latitude, longitude, number of bedrooms, number of bathrooms, square feet, and sales price.
We considered a model that predicts the selling price from the numbers of bedrooms, bathrooms, and square feet.
The data have some missing values, with 17\% of the samples missing at least one of the above three features.
We used the remaining 814 samples after excluding those with missing values.
We determined $\mathcal{E}$ using the latitude and longitude coordinates of each house.
The $\mathcal{E}$ is the set of subscript pairs $(i, j)$ corresponding to the five houses closest to the $i$-th house ($i=1,2,\cdots,n$) after removing the test set.

We considered the same three methods as in Section \ref{simulationStudy}: LL, BMN, and DLBMN.
The methods were compared using five-fold cross-validation.
Accuracies of each method were evaluated using prediction squared error (PSE).
The PSE corresponding to the $k$-th iteration of cross-validation is given by
\[
\mbox{PSE} =
\frac{1}{n^{(k)}}
\|\widetilde{\vector{y}}^{(k)}-\widetilde{X}^{(k)}\hat{\vector{\beta}}^{(k)}\|_2^2
,
\qquad
(k=1,2,\cdots,5)
,
\]
where $\widetilde{\vector{y}}^{(k)}$ and $\widetilde{X}^{(k)}$ are the test data in the cross-validation, $n^{(k)}$ is the number of test samples, and $\hat{\vector{\beta}}^{(k)}$ is the vector of the regression coefficient estimated by the data other than the test data.
We used the value of the hyperparameter that maximized the PSE.

All methods had a mean value and standard deviation of PSE of 0.37 and 0.09, respectively.
We defined $\mathcal{E}$ by the distances between houses.
It is assumed that the locations of properties and groups of individuals are highly related.
Therefore, the accuracy of relational coefficients in estimating LL and BMN is expected to be high.
In such a case, LL is known to estimate the model with high accuracy.
We confirm by this study that DLBMN enjoys the same accuracy as LL, even in environments where LL can estimate with high accuracy.

\section{Conclusion}

We proposed an approach to multi-task learning with network lasso based on Bayesian sparse modeling.
By grouping samples based on a Dirichlet--Laplace prior, we modeled the relationships among samples, which have not been considered by existing methods.
Compared with the existing methods, the proposed method estimated the parameters with high accuracy, regardless of the estimation accuracy of the relationships among the samples.
This result suggests that the fields of applicable data can be expanded compared to the existing methods.

\section*{Acknowledgements}
S. K. was supported by JSPS KAKENHI Grant Numbers JP19K11854 and JP20H02227.
Super-computing resources were provided by Human Genome Center (The Univ. of Tokyo).

\appendix
\def\thesection{Appendix \Alph{section}}

\section{Estimation by Gibbs sampling \label{gibbs_sammpling}}

For estimation by Gibbs sampling, it is necessary to derive the full conditional distribution from the joint prior distribution.
In the proposed model, the full conditional distributions are obtained from the prior distributions of each parameter and the likelihood function as follows:
\begin{eqnarray*}
  \vector{w}_i|\vector{y}, \{\tau_{i_1i_2}\}, \{r_{i_1i_2}\}, \{\widetilde{\tau}_{ij}\}, \sigma^2
  &\sim&
    \mbox{N}(S^{-1}\vector{m}, \sigma^2S^{-1}),
    \\
    &&
    S =
    \vector{x}_i\vector{x}_i^T+
    \lambda_1^2
    \sum_{k\in\mathcal{E}_i}
    \frac{r_{ik}^2}{\tau_{ik}}I_n+\widetilde{D}_i,
    \\
    &&
    \mathcal{E}_i =
    \{j ;\ (i, j) \in \mathcal{E}\},\quad
    \widetilde{D}_i =
    \mbox{diag}(\widetilde{\tau}^{-1}_{i1},\cdots,\widetilde{\tau}^{-1}_{ip}),
    \\
    &&
    \vector{m} =
    y_i\vector{x}_i
    +
    \lambda_1^2
    \sum_{k\in\mathcal{E}_i}
    \frac{r^2_{ik}}{\tau_{ik}}
    \vector{w}_k,
    \\
  \tau_{i_1i_2}^{-1}|W, r_{i_1i_2}, \sigma^2
  &\sim&
    \mbox{IGauss} (\mu', \lambda'),
    \\
    &&
    \mu' =
    \sqrt{
      \frac{\sigma^2}{\lambda_1^2r_{i_1i_2}^2\|\vector{w}_{i_1}-\vector{w}_{i_2}\|_2^2}
    },\quad
    \lambda' = 1,
    \\
  \lambda_1^{-1}|W, \{r_{i_1i_2}\}, \sigma^2
    &\sim&
    \mbox{giG} \left( \chi', \rho', \lambda' \right),
    \\
    &&
    \chi' = 2\sum_{(i_1,i_2)\in \mathcal{E}} r_{i_1i_2}
    \|\vector{w}_{i_1}-\vector{w}_{i_2}\|_2/\sqrt{\sigma^2},
    \\
    &&
    \rho' = 1, \quad
    \lambda' = \#\mathcal{E}(a-1),
    \\
  T_{i_1i_2} | W, \sigma^2
    &\sim& \mbox{giG}(\chi'' ,\rho'' ,\lambda''),
    \\
    &&
    \chi''= 2\|\vector{w}_{i_1}-\vector{w}_{i_2}\|_2/\sqrt{\sigma^2},
    \\
    &&
    \rho'' = 1, \quad
    \lambda'' = a-1,
    \\
    r_{i_1i_2}^{-1}
      &=&
      T_{i_1i_2} / \sum_{(i_1,i_2)\in \mathcal{E}} T_{i_1i_2},
    \\
  \widetilde{\tau}_{ij}^{-1}|W, \sigma^2
  &\sim&
    \mbox{IGauss} (\widetilde{\mu}', \widetilde{\lambda}'),
    \\
    &&
    \widetilde{\mu}' =
    \sqrt{ \frac{\lambda_2^2\sigma^2}{w_{ij}^2} }, \quad
    \widetilde{\lambda}' =
    \lambda_2^2,
    \\
  \sigma^2 | \vector{y}, W, \{\tau_{i_1i_2}\}, \{r_{i_1i_2}\}, \{\widetilde{\tau}_{ij}\}
    &\sim&
    \mbox{IG}(\nu', \eta'),
    \\
    &&
    \nu' = n + \#\mathcal{E} + np + \nu_0,
    \\
    &&
    \eta' =
    \sum_{i=1}^n(y_i-\vector{w}_i^T\vector{x}_i)^2 +
    \sum_{(i_1,i_2)\in\mathcal{E}}
    \frac{\lambda_1^2r_{i_1i_2}^2}{\tau_{i_1i_2}}
    \|\vector{w}_{i_1}-\vector{w}_{i_1}\|_2^2 \\
    &&\qquad
    +\sum_{i=1}^n\sum_{j=1}^{p}
    \frac{w_{ij}^2}{\widetilde{\tau}_{ij}} +
    \eta_0
    .
\end{eqnarray*}
These full conditional distributions make it possible for us to generate samples from the posterior distribution by Gibbs sampling.

\section{Dirichlet--Laplace prior distribution \label{dirichlet_laplace_prior}}

The Dirichlet--Laplace prior distribution \citep{bhattacharya2015dirichlet} was proposed to satisfy posterior consistency.
Bayesian regression models using this prior distribution are known to be asymptotically consistent in variable selection.
The probability density function of the Dirichlet--Laplace prior distribution is defined as follows:
\begin{eqnarray*}
  \mbox{DL}(\vector{\theta}|\alpha)
  &\propto&
  \int\cdots\int
  \prod_{j=1}^p
  \left\{
    p(\theta_j|\tau_j,\nu)
  \right\}
  p(\vector{\tau}|\alpha)
  p(\nu)
  \prod_{j=1}^p(d\tau_j)
  d\nu
  \\
  &\propto&
  \int\cdots\int
  \prod_{j=1}^p
  \left\{
    p(\theta_j|\psi_j,\tau_j^2,\nu^2)p(\psi_j)
  \right\}
  p(\vector{\tau}|\alpha)
  p(\nu)
  \prod_{j=1}^p(d\tau_jd\psi_j)
  d\nu
  ,
\end{eqnarray*}
where $\vector{\theta} = (\theta_1,\theta_2,\cdots,\theta_p)^T$ and $\vector{\tau} = (\tau_1,\tau_2,\cdots,\tau_p)^T$.
The prior distribution of each parameter is assumed as
\begin{eqnarray*}
  \theta_j|\tau_j,\nu
  &\sim&
  \mbox{Laplace}(1/\tau_j\nu),
  \\
  \theta_j|\tau_j,\psi_j,\nu
  &\sim&
  \mbox{N}(0, \psi_j\tau_j^2\nu^2),
  \\
  \vector{\tau}
  &\sim&
  \mbox{Dir}(\alpha,\cdots,\alpha),
  \\
  \psi_j
  &\sim&
  \mbox{Exp}(1/2),
  \\
  \nu
  &\sim&
  \mbox{Ga}(p\alpha,1/2),
\end{eqnarray*}
where the hyperparameter $\alpha\ (>0)$ is a parameter that controls the degree of sparseness of $\theta_j$.

\bibliographystyle{apalike}
\bibliography{main}
\end{document}